# Multi-Tier Tournaments: Matching and Scoring Players

Steven J. Brams
Department of Politics
New York University
New York, NY  10012
USA
steven.brams@nyu.edu

Mehmet S. Ismail
Department of Political Economy
King's College London
London WC2R 2LS
UK
mehmet.ismail@kcl.ac.uk

## Abstract

We introduce a novel system of matching and scoring players in tournaments, called Multi-Tier Tournaments, illustrated by chess and based on the following rules:

1. Players are divided into skill-based tiers, based on their Elo ratings.
2. Starting with one or more mini-tournaments of the least skilled players (Tier 1), the winner or winners— after playing multiple opponents—move to the next-higher tier.
3. The winners progress to a final tier of the best-performing players from lower tiers as well as players with the highest Elo ratings.
4. Performance in each tier is given by a player's Tournament Score (TS), which depends on his/her wins, losses, and draws (not on his/her Elo rating).

Whereas a player's Elo rating determines in which mini-tournament he/she starts play, TS and its associated tie-breaking rules determine whether a player moves up to higher tiers and, in the final mini-tournament, wins the tournament. This combination of players' past Elo ratings and current TS's provides a fair and accurate measure of a player's standing among the players in the tournament. We apply a variation of Multi-Tier Tournaments to the top 20 active chess players in the world (as of February 2024). Using a dataset of 1209 head-to-head games, we illustrate the viability of giving lower-rated players the opportunity to progress and challenge higher-rated players. We also briefly discuss the application of Multi-Tier Tournaments to baseball, soccer, and other sports that emphasize physical rather than mental skills.

*Keywords*: tournaments; schedule of matches; scoring of players; chess; games and sports

**1. Introduction**

There are many different kinds of tournaments, but here we focus on chess to illustrate our analysis, providing precise rules for matching and scoring players in a tournament. Later we discuss other kinds of tournaments to show the applicability of our analysis to different games of mental skill as well as sports like baseball and soccer that emphasize physical skills.

We introduce a novel system of matching and scoring players in tournaments, called Multi-Tier Tournaments, illustrated by chess and applying the following rules:

1. Players are divided into skill-based tiers, based on their Elo ratings.[1]
2. Starting with one or more mini-tournaments of the least skilled players (Tier 1), the winner or winners—after playing multiple opponents—move to the next-higher tier.
3. The winners progress to a final tier of the best-performing players from lower tiers as well as players with the highest Elo ratings.
4. Performance in each tier is given by a player's Tournament Score (TS), which depends on his/her wins, losses, and draws (not on his/her Elo rating).

Whereas a player's Elo rating determines in which mini-tournament he/she starts play, TS and its associated tie-breaking rules determine whether a player moves up to higher tiers and, in the final mini-tournament, wins the tournament. This combination of players' past Elo ratings and current TS's provides a fair and accurate measure of the player's standing among the players in the tournament. We apply a variation of Multi-Tier Tournaments to the top 20 active chess players in the world (as of February 2024). Using a dataset of 1209 head-to-head games that we collected, we illustrate the viability of giving lower-rated players the opportunity to progress and challenge higher-rated players. We also discuss the application of Multi-Tier Tournaments to baseball, soccer, and other sports that emphasize physical rather than mental skills.

---

[1] Elo (1978) ratings are a predictor of a match between two players, based on the difference in the players' ratings. Thus, when a low-rated player defeats a high-rated player, the Elo score of the low-rated player significantly increases and the high-rated player's score significantly decreases; on the other hand, the more similar the players' ratings are, the less the players' ratings are affected by the outcome of their match. In this manner, Elo ratings are self-correcting, providing a measure of the strength of players in a pool, all of whose members play each other. For more details, see https://en.wikipedia.org/wiki/Elo_rating_system. Like Elo, the tournament score (TS) we propose in rule 4 depends on a player's wins, draws, and losses, but only for players in each mini-tournament (more on this later).

The International Chess Federation (FIDE) prescribes several rules for organizing tournaments, which are distinctive in two major respects: (i) they have a fixed number of rounds, which is significantly less than the number of participants; (ii) they match players against opponents with similar scores after each round. Although the complete list of rules is extensive, the following are especially relevant to our paper:

1. Two players do not play against each other more than once.

2. The difference between the number of Black and the number of White games played by each player is not greater than two.

3. No player plays the same color three times in a row.

Rule (1) promotes diversity and reduces the potential bias arising from repeated matchups of the same players. Rule (2) promotes "color fairness," which may be undermined by the first-mover advantage of White in chess if one player plays more games with this color.[2] However, rule (2) allows a player to have one extra White game. Our analysis in Section 3 indicates a significant bias favoring the players who had an extra White game. For example, 66% of the top 10 finishers in major Swiss tournaments over the past decade had an extra White game. Rule (3) addresses "psychological fairness," because consecutively playing with the same color may confer a psychological advantage (when playing with White) or disadvantage (when playing with Black).

---

[2] To mitigate this bias, Brams and Ismail (2021) proposed a "catch-up" rule to put White and Black more on a par.

As we will show, our approach not only satisfies rules (2) and (3), and other standards of fairness, but it also better balances the distribution of colors than the currently used Swiss system. However, because it allows players to move into higher and higher levels of competition, it is possible for two players who meet early in a tournament to be later opponents. We do not think this is a serious violation of fair play, because knock-out tournaments that satisfy (1)—by eliminating competitors after one loss—are not necessarily fair to players in which a player's lapse in one game can be more than made up for by wins in the other games that he/she plays. We believe it is multiple games, at each tier in a tournament, that should be the determinant of who stays in and who is eliminated.

What is distinctive about our approach is its simplicity. It does not depend on indicators of performance other than wins, losses, and draws in multiple contests; where draws are not allowed or possible, this indicator can be dropped.[3] This makes it applicable to very different games and sports and, thereby, a general way of ranking competitors and choosing winners in tournaments.

## 2. Selecting and Scoring Players in Mini-Tournaments and Determining Winners

### *(i) Selection Algorithm*

Before players can be matched and scored in a tournament, they must be selected for a mini-tournament. Players in the first mini-tournament of Multi-Tier Tournaments (Tier 1) are those who have the lowest Elo ratings. For example, if there are 20 players in a tournament, they might be the 8 players with the lowest ratings. Each of these players plays against the 7 others in Tier 1 and receives a Tier Score (TS), based on his/her wins, losses, and draws (more on this later, in which we relax the assumption that every player plays against every other player in a mini-tournament).

**Example 1**

---

[3] We propose, however, rules for breaking ties when players have identical TS's. These are specific to chess and will need to be revised in applying Multi-Tier Tournaments to other games and sports.

To illustrate how players move up from an 8-player mini-tournament in a 20-player tournament, assume that the 2 players with the highest TS's in Tier 1 move up to Tier 2, joining the 6 players with the next-highest Elo scores. (Like Tier 1, Tier 2 would now have 8 players.) At the conclusion of the Tier 2 mini-tournament, the 2 players with the highest TS's in Tier 2 would move up to Tier 3, joining the 6 players with the highest Elo scores, so now Tier 3 also has 8 players. The Tier 3 players then play a third and final mini-tournament, the winner of which becomes the overall winner of the tournament.

Altogether, $8 + 6 + 6 = 20$ different players participate in up to three rounds of play. In Tier 1, each of the 8 players plays against the 7 other players, and the 2 players with the highest TS's move up to Tier 2. Next, each of the 8 Tier 2 players plays against the 7 other players in Tier 2, and the 2 players with the highest TS's move up to Tier 3. These 8 players play a final mini-tournament, and the player with the highest TS wins the tournament.

Players from Tier 1 who move up to Tier 2 play a total of $7+7 = 14$ games. If they succeed in Tier 2 and move up to Tier 3, they play a total of $7+7+7 = 21$ games. But if a player is not one of the top two in either Tier 1 or Tier 2, he/she is knocked out of the final competition.

The 6 players with the highest Elo scores, who start in Tier 3, play only 7 games, whether they win or lose in this mini-tournament. Only the single player with the highest TS in Tier 3 wins the tournament. Thus, a player starting out with a low or even moderate Elo rating, and therefore from a lower tier, means that he/she must play, and perform better, in more games than players who start out in the highest tier.

*(ii)  Scoring Algorithm and Breaking Ties*

In the mini-tournaments of each tier, each player *i* receives a Tier Score (TS$_{iJ}$) against all its opponents in subset *J* of his/her tier (as we will illustrate shortly, this may not be all members of his/her tier), according to the following formula:

$$\text{TS}_{iJ} = \frac{\text{\# of wins} - \text{\# of losses}}{\text{\# games played in tier}}$$

7$TS_{iJ}$ varies from -1 (for a player who loses all his/her games) to +1 (for a player wins all his/her games).

$TS_{iJ}$ differs from Elo ratings in not being a dynamically changing measure of player $i$'s strength, which may go up, down, or stay the same after each of a player's matches. Instead, $TS_{iJ}$ is a summary measure of $i$'s strength against his/her opponents in each of the mini-tournaments in which he/she plays (subscript *J in* $TS_{iJ}$ is the subset of these players). For example, a player could win a mini-tournament and move up to the next-higher tier, but in the latter mini-tournament he/she might do poorly and be eliminated from the tournament.

Notice that the denominator of $TS_{iJ}$ includes draws as well as wins and losses; the greater the number of drawn games, the less the absolute value of $TS_{iJ}$. If the numerator is positive, this lowers the value of $TS_{iJ}$ (more drawn games make $TS_{iJ}$ less positive). If the numerator is negative, this raises the value of $TS_{iJ}$ (more drawn games make $TS_{iJ}$ less negative). Thus, more drawn games hurt a winning player but help a losing player, as one would expect. If the numerator is 0, the number of drawn games is irrelevant.

Put another way, winning a game increases $TS_{iJ}$ more than drawing a game, and drawing a game increases TS more than losing a game. We think this is what a fair-scoring algorithm should do.

But what if the number of players in a mini-tournament is too great for each player to play a game against every other player in his/her starting tier? This will be the case when the number of players in a mini-tournament is greater than about 10, in which case we assume that players play against only a proper subset of opponents.[4] Before addressing the question of choosing such a subset, we suggest four rules to break ties when two or more players have the same $TS_{iJ}$'s at the end of a mini-tournament.

---

[4] One could add more tiers to solve this problem, but this may make the tournament unduly lengthy if each tier takes a few days to complete. Thus, if there are 10 players in a mini-tournament, and each player plays against 3 opponents a day, the mini-tournament will take three days to complete against the 9 opponents of each player.

When two players, A and B, have identical TS$_{iJ}$'s after play concludes in any tier, we propose the following rules to break ties:

(i) If A and B play each other in a mini-tournament and A defeats B, A is ranked above B. If they draw or do not play a game against each other, go to step (ii).
(ii) If A wins more games than B, A is ranked above B. (*Note*: B may also lose more games than A, but if B has more draws, it is possible that A and B may have the same TS$_{iJ}$.)
(iii) If A and B win the same number of games but A takes an average of fewer moves to defeat his/her opponents, A is ranked above B.
(iv) If (i), (ii), or (iii), in that order, fail to break a tie in TS$_{iJ}$'s between A and B—but their tie must be broken to determine which players go onto the next-higher tier or which player wins the tournament—use a random device to break the tie (e.g., by tossing a coin).

We think (i), (ii), and (iii) will break almost all ties between two players in a mini-tournament.[5] These tie-breaking rules, except for (iv), are consistent with the better performance of A or B. They can be extended to break ties among more than two players, and thereby to determine, if necessary, which player or players remain in the competition, move to a higher tier, or win the final mini-tournament and, therefore, the tournament.

**3. Constructing Fair Subsets of Players and Practical Concerns**

*(i)    Subset-Construction Algorithm*

If there are too many players in a tier for each player to play against all other players in that tier, then we need a way to find subsets of players whose average Elo rating is approximately the same. Thereby the level of competition that every player faces in his/her subset will be approximately the same as the level he/she would face in any other subset of opponents.

---

[5] Tie-breaking rule (iii) is perhaps least transparent. But we think defeating an opponent more quickly is consistent with being a stronger player.

**Example 2**

To illustrate how to construct equally competitive subsets, assume a 2-tier system, but now with 30 players in the lowest tier (Tier 1). For each player to play against his/her 29 opponents would take an inordinate amount of time. Instead, we propose that the 30 players be split into three subsets of 10 players each in such a way that each subset has about the same average Elo score.

To illustrate how to do this, assume that the top three players in Tier 1 (i.e., the winners of each of the three mini-tournaments in this tier) join 7 players in Tier 2 in a second and final mini-tournament comprising 7+3 = 10 players. Altogether, there are 37 players:

• *Tier 1 winners*: 3 winners of each of the Tier 1 subsets, who play 9 + 9 = 18 games in Tier 1 and Tier 2.

• *Tier 1 losers*: 30 – 3 = 27 players who play 9 games and are not the winners of their Tier 1 subsets.

• *Overall winner*: Winner of the Tier 2 mini-tournament, who may be either an original Tier 2 player, who plays 9 games, or one of the Tier 1 winners, who plays 9+9 = 18 games.

Each player in Tier 1 has the same opportunity to win in this tier and move up to Tier 2, whichever subset he/she is placed in, according to the following calculations for determining the composition of each of the three Tier 1 subsets of 10 players each:

1. For the $\binom{30}{10} = \frac{30!}{10!20!} = 30,045,015$ combinations (subsets) of 10 players each, determine the single subset whose 10 members come closest to having the same average Elo score as that of all 30 Tier 1 players (ties can be broken randomly). Thereby we identify which of the approximately 30 million subsets of 10 players each is closest to having the same average Elo score as that of all Tier 1 players (ties can be broken randomly):[6]

2. With this subset of 10 players determined, for the remaining $\binom{20}{10} = \frac{20!}{10!10!} =$ 184,756 combinations (subsets), identify the single subset whose 10 members come closest to having the same average Elo score as all 30 Tier 1 players.

3. Because the first two subsets each has about the same average Elo score as all 30 Tier 1 players, its last remaining subset of 10 players must also have about the same average Elo score.

Thereby we identify three Tier 1 subsets of 10 players that each have approximately the same average $TS_{iJ}$'s. Hence, our matching algorithm precludes any Tier 1 player from feeling discriminated against because he/she faces a higher level of competition in his/her 9 matches than would occur if he/she were in a different Tier 1 subset.[7]

Recall that we did not need the subset-selection algorithm in Example 1, because in each of the three tiers there were only 8 players, so the players in each tier could play against all others in the tier. In Example 2, this was not true for the 30 players in Tier 1, so we showed how to construct three homogeneous subsets of 10 players each that were equally competitive. Elo scores of players were used for this purpose, and they could be used again if too many players remain in contention in higher tiers to permit each player to play against every other.

---

[6] To be sure, 30 million is a large number but not beyond the ability of present-day computers to determine which subset(s) come closest to having the same average Elo rating as all Tier 1 players.

[7] This is a slight exaggeration, illustrated by the subset in which the player (say, A) with the highest Tier 1 Elo score appears. A will face a bit lower average level of competition than any player in the other two 10-member subsets, because A's Elo score pushes down the average score of the other 9 members of his/her subset. But if this player has a bit of a competitive advantage in his/her subset than players with lower Elo scores do in other subsets, this seems appropriate because they are the top players in their subsets, just as Tier 2 players enjoy an advantage by immediately being placed in Tier 2.

But it is TS's that determine who advances in a tournament. We believe that Elo scores should be used only to determine who, initially, gets placed in what tiers and—if the numbers in a higher-level tier are too great to permit each player to play against every other—to construct equally competitive subsets in that tier. In summary, an advancement from these subsets depends only on TS's, not Elo scores.

It is worth noting some concerns that real-life tournaments have raised. Perfect color fairness—whereby each player plays an equal number of games as Black and as White—is not always achievable, especially in tournaments with an odd number of rounds.

Even in tournaments with an even number of rounds, color fairness is not always satisfiable. For instance, in the 2023 FIDE Grand Swiss, which is one of the most prestigious Swiss chess tournaments, 9 of the top 10 finishers played one more game as White than as Black (6 vs. 5) (https://chess-results.com/tnr793016.aspx). A similar pattern was observed in the 2023 FIDE Women's Grand Swiss tournament (https://chess-results.com/tnr793017.aspx).

To see the effects of color imbalance on the outcomes in top-level Swiss tournaments, we analyzed the Grand Swiss and Grand Prix events organized by FIDE between 2017 and 2023.[8] These tournaments are of utmost importance in order to qualify for the Candidates Tournament, which determines the challenger for the world championship. We find a significant bias favoring the players who had an extra White game. Of the seven players who qualified for the Candidates Tournament from these events, six (86%) played at least one more game as White. Moreover, 66% of the top 10 finishers in all tournaments had an extra White game. These findings are consistent with White's historically higher win percentage in elite events, as reported in Brams and Ismail (2021).

Another concern of Swiss tournaments has been the varying strength of players, based on their Elo scores, in matches. In some cases, losing in the first-round game can lead to weaker opponents in subsequent rounds. The winner of the FIDE Grand Swiss 2023 not only had an additional game as White but also faced the weakest average opposition among the top five players, in part due to losing his first-round game (https://chess-results.com/tnr793016.aspx). Although the eventual winner's performance was undoubtedly exceptional, such small advantages can add up to influence the selection of the top two players, who qualify for the Candidates Tournament.

---

[8] The tournaments include Grand Swiss tournaments in 2023, 2021, and 2019 and Grand Prix tournaments in Sharjah, Moscow, Geneva, and Palma in 2017. We excluded previous Grand Prix tournaments that used the knockout format. As above, we have collected the data from https://chess-results.com.

To what extent are Multi-Tier Tournaments vulnerable to manipulation? For example, can a player benefit by deliberately losing or drawing against an opponent in order to be paired against weaker opponents in later rounds? Such an attempt at selective pairing is known as the "Swiss Gambit," perhaps because it was perfected in Swiss tournaments.

In Multi-Tier Tournaments, because it is always preferable to have a greater Elo score in order to enter a Multi-Tier tournament at a higher level, players have no incentive to lower their Elo scores by losing or drawing. Neither do they have an incentive to deliberately lose in a mini-tournament, because this lowers their TS's and hurts their chances of moving to a higher tier and, possibly, winning the tournament.

Another form of manipulation is for a player to withdraw from a tournament after pairings are announced to avoid playing against a player whom he/she thinks might beat or draw against him/her. This tactic is obviated by the difficulty of being able to anticipate one's opponents, even in Tier 1, because this may depend on the subset to which the matching algorithm assigns each player.

One way of countering such tactics is to impose hefty penalties for withdrawing in the middle of a tournament. But fines alone might not deter a top player like Leiner Dominguez, who in 2023 quit after two draws in an open tournament in Spain, saying that "I've come to a point where I'm simply risking too much if I continue [and lose or draw against a lower-rated player]" (https://www.chess.com/news/view/dominguez-quits-sitges-wesley-so-firouzja-candidates).

In a Multi-Tier Tournament, a highly rated player like Dominguez is assured that he will start against players mostly at his level. However, there may still be uncertainty about Dominguez's specific opponents if subsets are chosen in the manner we described, rendering uncertain whom he will play against if there is more than one subset in his tier. Nevertheless, he can rest assured that his opponents will be at roughly his level or have proved themselves by advancing to a high level in the tournament.

Another issue is the timing of matches. Under the Swiss system, matches for each round cannot be finalized until all games in the current round are completed. Because chess games are often lengthy, players may remain uninformed about their next opponents until late the evening before they play.

This uncertainty impacts players' ability to prepare. Players who can afford extensive support teams, including seconds and coaches, are at an advantage compared to those who must manage their preparation themselves.[9] Furthermore, this lack of early pairing information hinders television and online coverage, because broadcasters are unable to plan in advance for broadcasting key matchups that can attract fan viewership.

Our approach offers the following benefits:

1. Perfect or near-perfect White-Black color balance: If there is an even number of rounds, all players play Black and White the same number of times; if an odd number, all players play one color no more than one more round than the other.

2. Grouping of players into tiers with the most similar Elo ratings.

3. Prevention of two consecutive games with the same color.

We think our rules for matching and scoring players ensure a fair and comprehensive assessment of the players' performance in each mini-tournament, independent of the players' Elo ratings in their tier. They ensure that almost all players can be strictly ranked, even if their $TS_{iJ}$'s are the same, so there will be almost no doubt about who the few players are who move up to compete at the next-higher tier. Similarly, there will almost aways be a single winner of the last mini-tournament and, therefore, of the entire tournament.

---

[9] In chess, a second is a strong player assisting a player mainly during a tournament, focusing on analyzing opponents, preparing opening strategies, and offering psychological support. A coach, on the other hand, is involved in long-term training, working on a player's overall game, teaching new concepts, and improving a player's tactical and strategic understanding.

**4. Application of Multi-Tier Tournaments to Top 20 Chess Players**

In Table 1, we list the top 20 active chess players, based on their Elo ratings as of February 2024. As in Example 1, we divide players into three tiers: Initially, Tier 1 comprises the 8 players at ranks 1-8; Tier 2 comprises the 6 players at ranks 10-14; Tier 3 comprises the 6 players at ranks 15-20. Later, the two highest scorers in Tiers 1 and 2 move up, respectively, to Tiers 2 and 3, who then each have 8 members who compete against each other.

**Table 1. Elo Ratings of the Top 20 Players**

| Rank | Name | Rating | Rank | Name | Rating |
|---|---|---|---|---|---|
| 1 | M. Carlsen | 2830 | 11 | R. Praggnanandhaa | 2747 |
| 2 | F. Caruana | 2804 | 12 | Vidit G. | 2747 |
| 3 | H. Nakamura | 2788 | 13 | N. Abdusattorov | 2744 |
| 4 | Ding L. | 2762 | 14 | Gukesh D. | 2743 |
| 5 | A. Giri | 2762 | 15 | V. Keymer | 2738 |
| 6 | A. Firouzja | 2760 | 16 | A. Erigaisi | 2738 |
| 7 | I. Nepomniachtchi | 2758 | 17 | M. Vachier-Lagrave (MVL) | 2732 |
| 8 | W. So | 2757 | 18 | J.-K. Duda | 2732 |
| 9 | Wei Y. | 2755 | 19 | L. Aronian | 2725 |
| 10 | L. Dominguez | 2752 | 20 | S. Mamedyarov | 2722 |

Like Example 1, there is no need to construct fair subsets of players, because each of the 8 players in each tier can play against every one of the 7 other players in its tier to determine which 2 players move up to a higher tier. But unlike Example 1, the number of games played by each member of a tier against others in it is not 1, as in Multi-Tier Tournaments, but may vary radically, from 0 (there are seven pairs of players who play no games against each other) to 71, as in the case of the Carlsen (#1) - Aronian (#19) pair (Carlsen won 12, lost 8, and drew 51 games against Aronian).

It was not appropriate to calculate, for each tier, TS$_{iJ}$ of player *i* against all other players *J* in his/her/tier, because this measure is too much influenced by the outlier pairs in a tier, especially those pairs which happen to play many games against each other (perhaps because they are geographically proximate or are invited to, and choose to participate in, many tournaments). To remedy the problem of combining the wins, draws, and losses of such pairs with those pairs playing far fewer games, we define TA differently from how we did so in section 2.

For every two players, A and B, in a tier, we calculate A's TS in his/her tier as follows:

$$TS_{AB} = \frac{\text{\# of } A \text{ wins against } B \ - \ \text{\# of } B \text{ wins against } A}{\text{\# games played between } A \text{ and } B},$$

with $TS_{AB} = 0$ if *A* and *B* did not play any games. We illustrate the aggregation problem if we had used the formula for *TS* in section 2, *TS$_{iJ}$* (where subscript *J* indicates all opponents of *i*). Assume A plays a large number of games against B but many fewer games against the other 6 players, C, D, E, F, G, and H. In that case, the AB pair would have had an unduly large effect when we sum all the wins, draws and losses of A's opponents, including B, using *TS$_{iJ}$*.

But if we first normalize values of *TS* between -1 and +1 for each of A's opponents, as $TS_{AB}$ does, their summation does not give unduly large weight to the AB pair. Then, taking the average of *TS$_{AB}$* through *TS$_{AH}$*, we obtain a summary measure of how, hypothetically, A would do against all his/her opponents in his/her tier, but now based on varying numbers of games A has played against his/her opponents over 20 years (see next paragraph)—instead of just one game in a single tournament that never occurred.

**Table 2. A Three-Tiered Tournament with Top 20 Players Grouped into Three Tiers of 8, 6, and 6 Players, Based on Their Elo Ratings**[a]

| Rank | Tier 1 | $\overline{TS}$ | Tier 2 | $\overline{TS}$ | Tier 3 | $\overline{TS}$ |
|---|---|---|---|---|---|---|
| 1 | **MVL** | 0.192 | **MVL** | 0.18 | **Carlsen**[*] | 0.22 |
| 2 | **Aronian** | 0.191 | **Aronian** | 0.132 | Nakamura | 0.08 |
| 3 | Abdusattorov | 0.10 | So | 0.129 | Caruana | 0.04 |
| 4 | Mamedyarov | 0.041 | Wei | 0.00 | Aronian | -0.03 |
| 5 | Erigaisi | 0.040 | Dominguez | 0.00 | MVL | -0.04 |
| 6 | Gukesh | -0.01 | Nepomniachtchi | -0.07 | Ding | -0.06 |
| 7 | Duda | -0.10 | Praggnanandhaa | -0.17 | Giri | -0.07 |
| 8 | Keymer | -0.46 | Vidit | -0.20 | Firouzja | -0.13 |

[a]Two new players (highlighted in bold), who have the greatest average TS's in Tiers 1 and 2, are added first to Tier 2 and then to Tier 3. The TS winners of each tier are the two players ranked at the top of each tier, with Carlsen (starred), who starts in Tier 3, the winner of the tournament.

We used head-to-head contest data from 1209 games—excluding so-called rapid and blitz games that require somewhat different skills—that were played between May 2004 and February 2024 that is available from the following website that archives historical chess results (https://www.chessgames.com). Each matchup is based on an average of 14.4 games, with Tier 1 players averaging 6.6 games against each other, Tier 2 players averaging 10.1 games against each other, and Tier 3 players averaging 26.4 games against each other. Higher-ranked players tend to be older and are invited to more tournaments.[10]

As Table 2 shows, MVL and Aronian have the greatest TS's in both Tier 1 and 2. Hence, they advance to Tier 3. Consistent with Carlsen's dominance in chess for over ten years, he is the clear winner in this Multi-Tier Tournament, with an average TS of 0.22 in Tier 3. Nakamura and Caruana share the second and the third places, respectively. Notably, despite starting the competition in the lowest tier, MVL and Aronian perform better than three Tier 3 players.

---

[10] These are the average figures for each tier before winners from the lower tiers are added to the higher tiers.

## 5. Other Approaches to the Design of Fair Tournaments

FIDE has approved several versions of Swiss tournaments, including the Dutch, Burstein, Lim, and Dubov systems. These systems introduce changes to either the first-round pairing or later-round pairings under the Swiss system. For example, the Dubov system uses the average rating of the opponents to create pairings within the same score groups. However, this system still relies on the current score to match players and unlike our system, it does not necessarily improve the overall average of opponents' ratings or other desired aspects of matches (for a detailed overview, see Held, 2020). The most commonly used version is, in fact, the Dutch system, which matches players based on Elo ratings for the first round and then uses the players' updated scores for following rounds. For instance, in a six-player tournament with players ranked from 1 to 6, with 1 being the highest-rated, the first-round pairing would be 1–4, 2–5, and 3–6 according to the Dutch system.

While sharing similarities with league systems often used in sports like soccer, Multi-Tier Tournaments differ in several key aspects. First, unlike leagues, which run simultaneously over extended periods, Multi-Tier Tournaments progress sequentially within a single timeframe so that the players in the lowest tier have an opportunity to win the entire event. Second, leagues use a single or double round-robin format, whereas Multi-Tier Tournaments, similar to Swiss tournaments, are mainly designed for tournaments with a large number of players and a small number of rounds.

Recently, UEFA updated the format for the Champions League. Similar to both Swiss and Multi-Tier Tournaments, teams play fewer games than in a round-robin, and like Multi-Tier Tournaments, the format incorporates a degree of fairness by having teams play half their matches at home and half away. However, there are also key differences between this format and Multi-Tier Tournaments. Multi-Tier Tournaments use cardinal Elo ratings for fairer pairing allocation, while UEFA uses a complex set of criteria, including random allocation and ordinal rankings, to determine pairings. In addition, Multi-Tier Tournaments do not include a knockout component, whereas the UEFA Champions League does.

Various metrics to evaluate tournament success have been explored in the literature. Scarf et al. (2009) analyzed UEFA Champions League formats via simulations and found the round-robin to be the fairest in aligning pre- and post-tournament rankings. Similarly, Appleton (1995) and Sziklai et al. (2022) conducted simulations to assess different tournament formats, focusing on the win probability of the best teams and players. Key contributions in this field also include Glenn (1960) and Searls (1963), who both focus on the probability of the best players and teams winning in different tournament designs. As for Swiss tournaments, Olafsson (1990) explored algorithmic properties of different tournament designs, whereas Fuhrlich et al. (2021) and Sauer et al. (2024) proposed a Swiss-based pairing algorithm that produced improvements over in properties, like color fairness, over FIDE's Swiss systems. For a review of the tournament design literature, see the recent survey by Devriesere et al. (2024).

Tournament ranking systems have been theoretically explored by several researchers. Rubinstein (1980) axiomatized the standard points-based system, Csato (2017) studied Swiss system chess team tournaments, and Arlegi and Dimitrov (2020) examined fairness in knockout tournaments.

Tiebreaking in tournaments typically follows two approaches. One involves a final contest to resolve ties, as seen in tennis tiebreakers, soccer penalty shootouts, and some American football overtime games. These methods, and their fairness, have been the subject of several studies. Che and Hendershott (2008), Brams and Sanderson (2013), and Granot and Gerchak (2014) focus on bidding to determine which team picks among different options. Additionally, studies, for example, by Apesteguia and Palacios-Huerta (2010), Brams and Ismail (2018), Brams et al. (2018), Cohen-Zada et al. (2018), and Anbarci et al. (2021) focus on determining a fairer ordering of play in tiebreak contests.[11] The second approach analyses different rules for tiebreaking in Swiss chess tournaments (Anbarci and Ismail, 2024) and the FIFA World Cup group stage (Csato, 2021). For an overview of tiebreaking methods, see Csato (2023, Chapter 1.3).

The concept of strategyproofness, or incentive compatibility, in tournament rules has also been a subject of interest. Selected contributions in this area include works by Pauly (2014), Kendall and Lenten (2017), Brams et al. (2018), Dagaev and Sonin (2018), Csato (2019), Csato (2021), Anbarci et al. (2021), and Guyon (2022).

Finally, the literature on the maximum the number of consecutive games that can be played as White or Black in tournaments dates back to De Werra (1981), who studied round-robin soccer tournaments. In this context, playing at home is similar to playing as White, and playing away is similar to playing as Black. For a review of the relevant results, see Goossens and Spieksma (2012).

## 6. Conclusions

In contrast to Swiss tournaments, Multi-Tier Tournaments prioritize determining the pairings as fairly as possible prior to the start of games in a tier. Moreover, Multi-Tier Tournaments minimize, and eliminate whenever possible, biases related to the number of games played as White and Black (color fairness) and the number of consecutive games with each color (psychological fairness).

We showed how Multi-Tier Tournaments group players into tiers, and even within tiers, fairly. They also ensure that lower-rated players are afforded the opportunity to advance through the tiers, based on their performance. At the same time, they give a break to higher-rated players by allowing them to advance by playing fewer games.

Compared to knockout tournaments, where a single loss results in elimination, Multi-Tier Tournaments allow for a player's loss in one game to be countered by his/her winning in subsequent games. Although our application focused on chess, Multi-Tier Tournaments are applicable to other games and sports where Elo or other scoring methods are used, including tennis, soccer, and the big three of American sports—baseball, basketball, and football.

---

[11] For further empirical analysis of the first-mover advantage in penalty shootouts, see Kocher et al. (2012), Arrondel et al. (2019), Rudi et al. (2020), and Kassis et al. (2021).

In fact, one may consider a season of play in a sport like baseball a tournament of sorts, wherein teams play about the same number of games against teams in their league over the course of a season. Their records in these leagues determine which teams go into the playoffs, which might be considered mini-tournaments themselves, requiring winning, for example, 2 out of 3, 3 out of 5, or 4 out of 7 games.

In most sports leagues, the number of games that teams player over a regular season does not depend on how they are rated (it is different leagues that reflect different levels of skill). Only in the playoffs are the numbers narrowed down, as in Multi-Tier Tournaments, when the best-performing teams or players ascend to higher tiers as poorer teams are eliminated. Multi-Tier Tournaments show how this competition can be structured in a fair and systematic way.

444444

# References

Anbarci, Nejat, and Mehmet S Ismail. 2024. "AI-powered Mechanisms as Judges: Breaking Ties in Chess." *PLOS One*, forthcoming.

Anbarci, Nejat, Ching-Jen Sun, and M Utku Unver. 2021. "Designing practical and fair¨ sequential team contests: The case of penalty shootouts." *Games and Economic Behavior* 130:25–43.

Apesteguia, Jose, and Ignacio Palacios-Huerta. 2010. "Psychological pressure in competitive environments: Evidence from a randomized natural experiment." *American Economic Review* 100 (5): 2548–64.

Appleton, David R. 1995. "May the Best Man Win?" *Journal of the Royal Statistical Society Series D: The Statistician* 44 (4): 529–538.

Arlegi, Ritxar, and Dinko Dimitrov. 2020. "Fair elimination-type competitions." *European Journal of Operational Research* 287 (2): 528–535. https://doi.org/https://doi.org/10.1016/j.ejor.2020.03.025.

Arrondel, L., Duhautois, R., and Laslier, J. F. (2019). Decision under psychological pressure: The shooter's anxiety at the penalty kick. *Journal of Economic Psychology*, 70, 22–35.

Brams, Steven J, and Mehmet S Ismail. 2018. "Making the Rules of Sports Fairer." *SIAM Review* 60 (1): 181–202.

———. 2021. "Fairer Chess: A Reversal of Two Opening Moves in Chess Creates Balance Between White and Black." In *2021 IEEE Conference on Games (CoG),* 1–4. IEEE.

Brams, Steven J, Mehmet S Ismail, D Marc Kilgour, and Walter Stromquist. 2018. "Catch-Up: A Rule That Makes Service Sports More Competitive." *American Mathematical Monthly* 125 (9): 771–796.

Brams, Steven J., and Zachary N. Sanderson. 2013. "Why you shouldn't use a toss for overtime." *+Plus Magazine,* https://plus.maths.org/content/toss-overtime.

Che, Yeon-Koo, and Terrence Hendershott. 2008. "How to divide the possession of a football?" *Economics Letters* 99 (3): 561–565. https://doi.org/10.1016/j.econlet.2007.10.013

Cohen-Zada, Danny, Alex Krumer, and Offer Moshe Shapir. 2018. "Testing the effect of serve order in tennis tiebreak." *Journal of Economic Behavior & Organization* 146:106–115.

Csato, Laszlo. 2017. "On the ranking of a Swiss system chess team tournament." *Annals of Operations Research* 254 (1-2): 17–36.

———. 2019. "UEFA Champions League entry has not satisfied strategyproofness in three seasons." *Journal of Sports Economics* 20 (7): 975–981.

———. 2021. *Tournament Design: How Operations Research Can Improve Sports Rules.* Springer Nature.

———. 2023. "How to avoid uncompetitive games? The importance of tie-breaking rules." *European Journal of Operational Research* 307 (3): 1260–1269.

Dagaev, Dmitry, and Konstantin Sonin. 2018. "Winning by Losing: Incentive Incompatibility in Multiple Qualifiers." *Journal of Sports Economics* 19 (8): 1122–1146.

Devriesere, Karel, László Csató, and Dries Goossens. 2024. "Tournament design: A review from an operational research perspective." *arXiv preprint arXiv:2404.05034*.

De Werra, Dominique. 1981. Scheduling in sports. In Hansen, P., editor, *Studies on Graphs and Discrete Programming*, volume 11 of Annals of Discrete Mathematics, pages 381–395. North-Holland, Amsterdam, The Netherlands.

Elo, Arpad E. 1978. *The Rating of Chess Players, Past and Present*. New York: Arco Publishing.

Fuhrlich, Pascal, Agnes Cseh, and Pascal Lenzner. 2022. "Improving Ranking Quality and Fairness in Swiss-System Chess Tournaments." In *Proceedings of the 23rd ACM Conference on Economics and Computation,* 1101–1102. EC'22. Boulder, CO, USA: Association for Computing Machinery

Glenn, WA. 1960. "A Comparison of the Effectiveness of Tournaments." *Biometrika* 47 (3/4): 253–262.

Goossens, Dries R., and Spieksma, Frits C. 2012. Soccer schedules in Europe: An overview. *Journal of Scheduling*, *15*, 641–651.

Granot, Daniel, and Yigal Gerchak. 2014. "An auction with positive externality and possible application to overtime rules in football, soccer, and chess." *Operations Research Letters* 42 (1): 12–15. https://doi.org/https://doi.org/10.1016/j. orl.2013.11.001

Guyon, Julien. 2022. "'Choose your opponent': A new knockout design for hybrid tournaments." *Journal of Sports Analytics* 8 (1): 9–29.

Held, Mario, and Swiss Dubov. 2020. *"*A Comparison Between Swiss Pairing Systems.. https://spp.fide.com/wp-content/uploads/2020/11/Dubov-vs-FIDE-Swiss.pdf. Accessed: 2023-11-24.

Kassis, Mark, Sascha L. Schmidt, Dominik Schreyer, and Matthias Sutter. 2021. "Psychological Pressure and the Right to Determine the Moves in Dynamic Tournaments – Evidence from a Natural Field Experiment." *Games and Economic Behavior* 126 (March): 771–796.

Kendall, Graham, and Liam J.A. Lenten. 2017. "When sports rules go awry." *European Journal of Operational Research* 257 (2): 377–394. https://doi.org/https://doi.org/10.1016/j.ejor.2016.06.050.

Kocher, Martin G., Marc V. Lenz, and Matthias Sutter. 2012. "Psychological Pressure in Competitive Environments: New Evidence from Randomized Natural Experiments." *Management Science* 58 (8): 1585–91.

Lambers, Roel, and Frits C. R. Spieksma. 2021. "A Mathematical Analysis of Fairness in Shootouts." *IMA Journal of Management Mathematics* 32 (4): 411–24.

Olafsson, Snjolfur. 1990. "Weighted Matching in Chess Tournaments." *Journal of the Operational Research Society* 41 (1): 17–24.

Pauly, Marc. 2014. "Can strategizing in round-robin subtournaments be avoided?" *Social Choice and Welfare* 43 (1): 29–46.

Rubinstein, Ariel. 1980. "Ranking the participants in a tournament." *SIAM Journal on Applied Mathematics* 38 (1): 108–111.

Rudi, Nils, Marcelo Olivares, and Adity Shetty. 2020. "Ordering Sequential Competitions to Reduce Order Relevance: Soccer Penalty Shootouts." *PLOS One* 15 (12): 1–11.

Sauer, Pascal, Cseh, Ágnes and Lenzner, Pascal. 2024. "Improving ranking quality and fairness in Swiss-system chess tournaments." *Journal of Quantitative Analysis in Sports*, 20 (2): 127–146.

Scarf, Philip, Muhammad Mat Yusof, and Mark Bilbao. 2009. "A numerical study of designs for sporting contests." *European Journal of Operational Research* 198 (1): 190–198.

Searls, Donald T. 1963. "On the probability of winning with different tournament procedures." *Journal of the American Statistical Association* 58 (304): 1064–1081.

Sziklai, Balázs R., Péter Biró, and László Csató. 2022. "The efficacy of tournament designs." *Computers & Operations Research* 144: 105821.